\documentstyle[preprint,aps]{revtex}
\tighten
\begin{document}
\draft
\title{A Class of Fast Methods for Processing Irregularly Sampled\\
or Otherwise Inhomogeneous One-Dimensional Data}
\author{George B. Rybicki and William H. Press}
\address{Harvard-Smithsonian Center for Astrophysics, 60 Garden St.,
Cambridge, MA 02138}
\date{\today}
\maketitle
\begin{abstract}
With the {\em ansatz} that a data set's correlation matrix has a
certain parametrized form (one general enough, however, to allow the
arbitrary specification of a slowly-varying decorrelation distance and
population variance) the general machinery of Wiener or optimal
filtering can be reduced from $O(n^3)$ to $O(n)$ operations, where $n$
is the size of the data set.  The implied vast increases in computational
speed can allow many common sub-optimal or heuristic data analysis
methods to be replaced by fast, relatively sophisticated, statistical
algorithms.  Three examples are given: data rectification, high- or
low- pass filtering, and linear least squares fitting to a model with
unaligned data points.
\end{abstract}


\narrowtext

% body of paper here
In a previous analysis of irregularly spaced observations of the
gravitationally lensed quasar 0957+561 \cite{Paper1,Paper2} we have
used to good effect the full machinery of Wiener (or optimal)
filtering in the time domain, including the use, where appropriate, of
unbiased (``Gauss-Markov'') estimators \cite{Paper3}.  These, and
related, techniques are general enough to be applicable to data that
is not only irregularly sampled (including the notoriously common case
of ``gappy'' data), but also to data that is highly inhomogeneous in
its error bars, or with point-to-point correlations (so that errors
are not independent).  The principal reason that these methods are not
better known and more widely used seems to be the fact that their use
entails the numerical solution of sets of linear equations as large as
the full data set.  Only recently have fast workstations allowed
application to data sets as large as a few hundred points; sets larger
than a few thousand points are currently out of reach even on the
largest supercomputers, since the computational burden for $n$ data
points scales as $n^3$.  As an example, the analysis in \cite{Paper2}
(leading to a measurement of the offset in time of the two radio
images of the lensed quasar) required overnight runs on a fast
workstation.

In this context, we were therefore quite surprised recently to notice
that the introduction of a particular simplifying assumption
(essentially the {\em ansatz} of a certain parametrized form of the
data's correlation function) allows all the calculations already
mentioned, and many more, to be done in {\em linear} time, that is,
with only a handful of floating operations per data point.  In fact we
have verified that we are able to obtain results substantially
identical to \cite{Paper2} in less than 3 seconds of computer time,
about $10^4$ times faster than the previous analysis.

Speed increases of $10^4$ or greater (that is, from $O[n^3]$ to $O[n]$
for $n$ data points) are not merely computer time savers.  Such
increases are {\em enabling} for the application of sophisticated
statistical techniques to data sets that hitherto have been analysed
only by heuristic and ad-hoc methods.  The Fast Fourier Transform
(FFT) is a previous example of a numerical algorithm whose raw speed
caused it to engender a considerable universe of sophisticated
applications.  By their nature, FFT methods are generally not
applicable to irregularly sampled, or otherwise inhomogeneous, data
sets (though see \cite{PR89}).  Although the methods we describe here
are not related to the FFT in a mathematical sense, we think they have
the potential to be comparably significant in engendering new and
powerful techniques of data analysis.  In the interest of making such
new methods available to the widest possible community, we outline, in
this Letter, the mathematical foundation of the class, and give three
examples of early applications.  We will also make available, via the
Internet \cite{netnote}, a ``developer's kit'' of Fortran-90 code,
fully implementing the examples given here.

We begin with the observation that many, if not most, one-dimensional
processes of interest (e.g., measurements as a function of time $t$)
have a characteristic decorrelation time (which may itself vary with
time), so that a set of measurements $y_i$ at the ordered times $t_i$,
$i=1,\ldots,n$, have an expected (population) correlation matrix that
is peaked on the diagonal $i=j$ and decays away from the diagonal in
both directions.  We consider the case where this decay can be
modeled, even if only roughly, by the form
\begin{equation}
\Phi_{ij} \equiv
  \exp \left[ - \left| \int_{t_i}^{t_j} w(t) dt \right|\;\right]
\end{equation}
where $w(t)$, the reciprocal of the decorrelation time, can
be thought of a slowly varying with time (or constant).  All our results
derive from the remarkable fact that the inverse of the matrix (1),
$\Phi^{-1}_{ij}\equiv T_{ij}$, is tridiagonal with
\begin{equation}
T_{ij} =\cases{
1+r_1e_1 & $i=j=1$\cr
-e_i & $1<i=j-1<n-1$\cr
1+r_ie_i+r_{i-1}e_{i-1} &  $1<i=j<n$\cr
-e_j & $1<j=i-1<n-1$\cr
1+r_{n-1}e_{n-1} & $i=j=n$\cr
0 & otherwise\cr}
\end{equation}
where
\begin{equation}
 r_i \equiv \exp\left[-\left| \int_{t_i}^{t_{i+1}} w(t) dt
   \right|\;\right]
\end{equation}
and
\begin{equation}
 e_i \equiv (r_i^{-1}-r_i)^{-1}
\end{equation}

It is well known \cite{Golub89,NR} that tridiagonal systems can be
solved in linear time, requiring $8n$ arithmetic operations.  A first
surprising result is therefore that the operation of multiplying the
{\em non-sparse} matrix
$\Phi$ by any data vector $y$,
\begin{equation}
 I_i = \sum_j \Phi_{ij} y_j
\end{equation}
can be rendered fast by instead solving (for $I$) the tridiagonal problem
\begin{equation}
 \sum_j T_{ij} I_j = y_i
\end{equation}

The matrix $\Phi$ by itself is not a very good correlation matrix, since
it is normalized to unity on the diagonal.  This is easily remedied by
the introduction of another (generally slowly varying or constant)
function $V(t)$ to represent the population variance of the process, that
is, the typical or {\em a priori} mean square amplitude for a measurement
$y$ at time $t$.  Our correlation matrix {\em ansatz} is then
\begin{equation}
 C_{ij} = \left[ V(t_i)V(t_j) \right]^{1/2}
\Phi_{ij}
\end{equation}
which has the (tridiagonal) inverse
\begin{equation}
 [C^{-1}]_{ij} = \left[ V(t_i)V(t_j) \right]^{-1/2} T_{ij}
\end{equation}
(The function $V$ is not to be confused with the error or noise in a
single measurement, which need not be slowly varying, see below.)

In fact, equations (1)--(6) represent just one example that
derives from the more general
fast evaluation of ``forward plus backward'' integrals with the form
\begin{equation}
 I_i = F_i + B_i \qquad i=1,\ldots,n
\end{equation}
where the forward and backward pieces have individual exponentially
decaying recurrences,
\begin{eqnarray}
F_{i+1} &=& F_i r_i+f_{i}\nonumber\\
B_{i-1} &=& B_i r_{i-1}+b_{i-1}
\end{eqnarray}
where the $f_i$ (or $b_i$) is the forward (or backward) increment
across the single interval $(t_i,t_{i+1})$.  The key observation
is that $I_j$'s can be found
from the $f$'s and $b$'s (without the intermediate calculation of
$F$'s and $B$'s) by solving the tridiagonal system
\widetext
\begin{equation}
\sum_j T_{ij} I_j =
\cases{
F_1 + e_1(b_1/r_1-f_1) & $i=1$\cr
e_{i-1}(f_{i-1}/r_{i-1}-b_{i-1})+e_{i}(b_{i}/r_{i}-f_{i})
   & $2\leq i\leq n-1$\cr
e_{n-1}(f_{n-1}/r_{n-1}-b_{n-1})+B_n & $i=n$}
\end{equation}
\narrowtext
The special case of eq.~(6) is recovered by
\begin{equation}
 f_i = (y_{i+1}+r_{i}y_{i})/2 \qquad b_i = (y_i+r_i y_{i+1})/2
\end{equation}
However, other choices for $f_i$ and $b_i$, e.g., corresponding
to quadrature formulas across the interval $(t_i,t_{i+1})$, have other
uses, as we will see below.

This is all the machinery we need for the three examples that we now
give.  The idea that certain special correlation matrices can have
simple inverses is not new \cite{Roy1,Roy2,Graybill}, but we are not
aware of the previous use in data processing of tridiagonally fast
forms as general as eqs.~(1)--(4), (7)--(8), or (9)--(12).

{\em Example 1: Wiener filtering and data rectification.} Here, one is
given an irregularly sampled data set $y_i=y(t_i)$, $i=1,\ldots,n$,
with error estimates $\sigma_i$.  The $\sigma_i$'s may be highly
variable, with well-measured values intermixed with poorly-measured
ones.  One desires best estimates (in the sense, with some technical
assumptions, of minimum variance, see [3]) of the underlying signal
$s(t)$ either at the measured times $t_i$ (Wiener filtering), or else
at some different, usually equally spaced, set of times $t^\prime_j$,
$j=1,\ldots,m$ (data rectification).  In the real world, data
rectification is now often accomplished by linear interpolation
between nearest measured points.  Such interpolation can give highly
sub-optimal results, since it can use a poorly measured near point
instead of a much better measured point only negligibly farther away.
The procedure described here uses, in effect, an optimal combination
of points, weighting them appropriately by their combination of
nearness to the desired point and smallness of their measured error.

One proceeds as follows.  Step 1.1: estimate the inverse decorrelation
length $w(t)$ and the population variance $V(t)$.  These estimates
need not be very accurate, since the error in the result of Wiener
filtering is second order in any error in the filter.  It generally
suffices to use constant values $w$ and $V$.

Step 1.2: Form an ``augmented'' vector of times $t_i$, $i=1,\ldots,N$,
consisting of the union of the times at which data is measured and the
times at which output is desired.  (If there are no overlaps, then
$N=m+n$.)  Form a corresponding augmented data vector $\mbox{\bf y}_*$, with
measured data values in the appropriate slots, an arbitrary constant
value (generally the mean of the measured values) in the other slots.
Form an augmented ``reciprocal error'' vector by similarly combining
the the measured $1/\sigma_i$ values
(in the measured slots) with a value zero in the unmeasured slots.

Step 1.3: Calculate the output of the Wiener filter, by using the
right-hand form of the matrix equation
\begin{equation}
\widehat{{\bf s}} =
\mbox{\bf S}^T[\mbox{\bf S}+\mbox{\bf N}]^{-1} \mbox{\bf y}_*
= [\mbox{\bf N}^{-1}+\mbox{\bf S}^{-1}]^{-1} \mbox{\bf N}^{-1} \mbox{\bf y}_*
\end{equation}
Here $\mbox{\bf N}^{-1}$ is the diagonal matrix formed from the
square of the reciprocal error vector
$N^{-1}_{ij} = \left(1/\sigma_i^2 \right) \delta_{ij}$
while $\mbox{\bf S}$ is the correlation matrix
(called $\mbox{\bf C}$ in eq.~7, above),
fully specified by the sets $V_i$, $w_i$, $t_i$.  In particular,
since $\mbox{\bf S}^{-1}$ is tridiagonal, right-hand term in brackets is also
tridiagonal, and $\widehat{{\bf s}}$ can be obtained by a single fast
tridiagonal solution.

Step 1.4:  Unpack the desired results from $\widehat{{\bf s}}$, either
the values at the measured $t_i$'s, or the values at the rectified
$t^\prime_i$'s, or both.

{\em Example 2.  Low- or High-pass Filtering of Irregularly Spaced
Values.}  Here, for brevity, we assume that the data is error free.
(A more complicated example would combine the optimal estimation of
Example 1 with the filtering described here.) At first sight, a
matrix like eq.~(1) does not look like a very good low-pass filter,
because the discontinuity in slope on the diagonal introduces significant
high-frequency leakage.  An important trick, however, is to allow
$w$ to be complex, but take the filter to be the real part of the result.
Then, there is a {\em unique} exponential filter with all the following
properties: (i) no discontinuity in the derivative of the impulse
response, (ii) frequency response unity at $f=0$ and flat through
the third derivative, (iii) response falls off as $f^{-4}$ (amplitude),
or 24 dB per octave (power),
for $f>f_c$, where $f_c$ is the 3 dB cutoff frequency.  Using this
exponential, the filter can be derived from eq.~(11) by the application of
a simple quadrature rule to the intervals between the points.

Step 2.1:  Calculate the complex values
\begin{equation}
 W_j = K (1+i) f_c \left| t_{j+1}-t_j \right| \qquad j=1,\ldots,n-1
\end{equation}
where $K=5.53807$ for a low-pass filter, $K=3.56427$ for a high-pass filter.
(The different values simply scale the respective 3 dB points to $f_c$.)
For each $W_j$, calculate $r_j\equiv \exp[-W_j]$ (replacing eq.~3).

Step 2.2:  Solve the complex tridiagonal system
\widetext
\begin{equation}
 \sum_j T_{ij} u_j = {1\over 2} \times \cases{
(s_1-s_2)/W_1 & $i=1$\cr
(s_i-s_{i+1})/W_i + (s_i-s_{i-1})/W_{i-1} & $i=2,\ldots,n-1$\cr
(s_n-s_{n-1})/W_{n-1} & $i=n$\cr}
\end{equation}
where the $s_i$'s are the input values.
\narrowtext

Step 2.3: The outputs of the high- or low-pass filters, at the
locations of the input data, are given respectively by
\begin{equation}
 {\cal H}\mbox{\bf s}={\rm Re}(\mbox{\bf u}) \qquad
{\cal L} \mbox{\bf s}=\mbox{\bf s}-{\rm Re}(\mbox{\bf u})
\end{equation}
(It is worth remarking that, by a different choice of complex exponential,
one can similarly approximate Gaussian convolution and deconvolution.)

{\em Example 3.  Linear least squares fitting a model, where the data
points and model points are not available at the same positions.} This
is just one of many possible extensions of Example 1.  One desires the
coefficient vector $\mbox{\bf q}$ that best fits a linear combination of model
basis functions.

Step 3.1 is the same as Step 1.1, above.  Step 3.2 is similar to step
1.2.  Model (as opposed to data) slots should, however, be filled with
zero values for both $\mbox{\bf y}_*$ and $\mbox{\bf N}$.
Additionally form an augmented
matrix $\mbox{\bf L}_*$ each of whose columns contains the values of a single
model basis function (in rows corresponding to the model slots) or zero
(in rows corresponding to the data slots).

For Step 3.3, as in Step 1.3, note that the inverse
\begin{equation}
\mbox{\bf C}^{-1} \equiv (\mbox{\bf S}+\mbox{\bf N})^{-1} =
\mbox{\bf S}^{-1}[\mbox{\bf 1}+\mbox{\bf N}\mbox{\bf S}^{-1}]^{-1}
\end{equation}
can be calculated by a single tridiagonal solution, followed by
an (also fast) tridiagonal left multiplication by $\mbox{\bf S}^{-1}$.
Thus the coefficient matrix, which is given by
\begin{equation}
\mbox{\bf q} = [\mbox{\bf L}^T_*(\mbox{\bf C}^{-1}\mbox{\bf L}_*)]^{-1}
\mbox{\bf L}^T_*(\mbox{\bf C}^{-1}\mbox{\bf y}_*)
\end{equation}
can be calculated by performing fast operations on the columns
of $\mbox{\bf L}_*$ and on $\mbox{\bf y}_*$.

Step 3.4:  Using the inverses already calculated for $\mbox{\bf q}$, the
$\chi^2$ for the fit is calculated as
\begin{equation}
 \chi^2 =
(\mbox{\bf y}_*-\mbox{\bf L}_*\mbox{\bf q})^T [ (C^{-1}\mbox{\bf y}_*)
-(C^{-1}\mbox{\bf L}_*)\mbox{\bf q}]
\end{equation}
By repeating the calculations of eqs.~(18) and (19),
this quantity can be minimized with respect to any additional parameters
in the fit that enter nonlinearly.  (We have applied this procedure
with great success to a problem where the nonlinear parameter is a
time-scaling of the data, so that the positions of the measured data
points are shifted and compressed or expanded  with each iteration.
Use of  eqs.~(18) and (19) avoids the necessity of recalculating the
model at each iteration.)

The examples given here are only the simplest of many related techniques
that share the underpinnings of (i) non-sparse matrices with
tridiagonal (or, more generally, band-diagonal or other fast) inverses,
and (ii) judicious factorization to maintain fast
evaluation at each step, as in eqs.~(13) and (17).
Single exponential forms are only the simplest of a larger family.
For example, correlation matrices that are approximated by the sum
of {\em two} exponential forms with tridiagonal inverses can
be treated by the factorization
\begin{equation}
 (\mbox{\bf T}_1^{-1}+\mbox{\bf T}_2^{-1} + \mbox{\bf N})^{-1} =
\mbox{\bf T}_1(\mbox{\bf T}_2+\mbox{\bf T}_1
+\mbox{\bf T}_2\mbox{\bf N}\mbox{\bf T}_1)^{-1}\mbox{\bf T}_2
\end{equation}
where $\mbox{\bf T}_1$ and $\mbox{\bf T}_2$ are each tridiagonal,
$\mbox{\bf N}$ is diagonal.
The term in parentheses on the right is pentadiagonal, therefore
also admitting fast inversion.  Although the equations become more
complicated, a sum of $k$ exponential forms can be cast as a band-diagonal
inversion with bandwith $2k+1$.

This work was suppported in part by the National Science Foundation
(PHY-91-06678).


\begin{references}
\bibitem{netnote} Code can be obtained by anonymous ftp to
cfata4.harvard.edu.  The ascii file to be downloaded is
``pub/fast-method-toolkit''.
\end{references}
\end{document}